\documentclass[10pt, conference, compsocconf]{IEEEtran} 
\usepackage{graphicx}
\usepackage{framed}
\usepackage{fancyvrb}
\IEEEoverridecommandlockouts
\ifCLASSINFOpdf
\else
\fi
\begin{document}
%
\title{Experimenting with Component-Based Middleware for \\Adaptive Fault Tolerant Computing$^{\dag}$\thanks{$\dag$This work has been supported by the French National Research Agency ANR, contract nr.  ANR-BLAN-SIMI10-LS-100618-6-01
}\vspace{-.3cm}}


\author{\IEEEauthorblockN{Miruna Stoicescu ~ ~ Jean-Charles Fabre ~ ~ Matthieu Roy}
\IEEEauthorblockA{LAAS-CNRS\\
Universit\'e de Toulouse\\
Toulouse, France}}
 
\maketitle

\begin{abstract}
This short paper describes early experiments to validate the capabilities of a component-based platform to observe and control a software architecture in the small. This is part of a whole process for resilient computing, i.e. targeting the adaptation of fault-tolerance mechanisms at runtime.
\end{abstract}


\IEEEpeerreviewmaketitle

\section{Problem statement}
\label{problem}
Dependable systems designers must
ensure that dependability properties are not violated by
the target system evolution. 
This leads to the notion of resilient computing \cite{laprie2008dependability}. Dependable systems consist of a functional layer to which are associated one or several fault tolerance mechanisms (FTMs). The choice of an appropriate FTM is based on several criteria: fault model, application assumptions such as determinism and state accessibility, available resources.

Should the evolution of the system during its operational lifetime entail one or several changes in the values of these parameters, the initial FTM could become useless and, more importantly, inappropriate wrt the actual fault model that may evolve during system's lifetime. A complete restart with new FTMs is not a solution for systems which must not stop for a long period of time. Our work aims at enabling the adaptation of FTMs through a differential approach minimizing the modifications to perform a transition from one FTM to another one. 

To reach our goal, we need software engineering tools enabling modular design and exploring and manipulating the software architecture at runtime, namely component-based middleware.This paper reports on the design and the experimental component-based implementation of a FTM, a Primary-Backup Replication (PBR) and its manipulation at runtime.
 
\section{Component-based Design}
\label{design}
Service Component Architecture (SCA) provides a set of specifications for building and composing loosely-coupled, tailorable applications encompassing a wide range of technologies. The main idea of this paradigm is that applications are built from {\em bricks} (i.e., components) exposing their functionalities in the form of services. Such an approach facilitates reuse and evolutivity as components consume services provided by other components without being aware of how they are implemented.

The component-based middleware on which we develop our  mechanisms is OW2 FraSCAti~\cite{frascati}, a platform providing runtime support for SCA. FraSCAti offers support for runtime reconfigurations of the application's architecture in several ways, one of which being {\tt FScript}~\cite{fscript}, a script language for exploring and modifying component-based systems. This  support is necessary to fulfill the requirements we identified for building a resilient computing framework:
\begin{itemize}
\item {\em access} to components' state and properties;
\item {\em control} over components' {\em lifecycle} ({\tt start}, {\tt stop});
\item {\em control over interactions} between components, i.e. for destroying or creating bindings.
\end{itemize}

\section{Current Implementation}

\begin{figure}[htbp]
\centering
\includegraphics[scale=0.45]{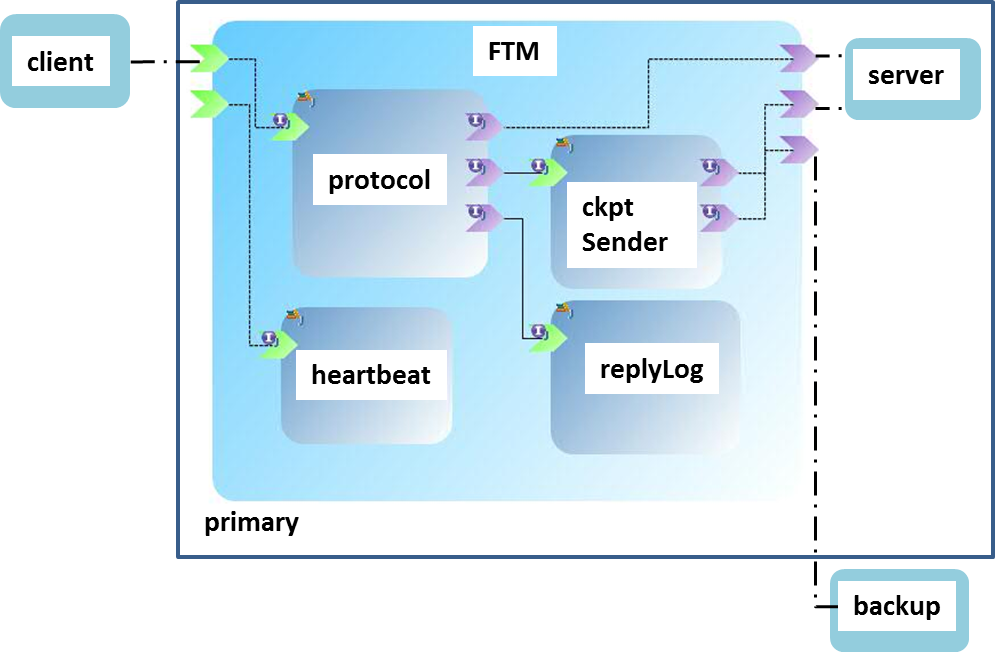}
\caption{PBR (Primary-Backup Replication) component-based design}
\label{pbr}
\end{figure}

Figure \ref{pbr} shows the component-based design of PBR, with a detailed view of the primary entity. All components are implemented in Java. Our application consists of three components and their interactions: the \emph{client}, the \emph{primary} processing the requests, the \emph{backup} processing the checkpoints sent by the primary. In parallel, a failure detector periodically checks liveness of the primary using a heartbeat mechanism. Should the primary crash, the backup substitutes for it.

The component-based design emphasizes the separation of concerns between the functional layer of the application (the actual server) and the non-functional one (the FTM protocol, primary-side and backup-side). This kind of separation, together with a careful design of the FTM supported by the freedom to build components as fine-grained as we want, give us a strong degree of control over the architecture of our application. We exercise this control at runtime for manipulating the components by modifying the values of their properties, their connections, stopping, restarting, removing, replacing them, etc.

\section{Experimenting with dynamics at runtime}
In this first experiment, the objective is not the to fully adapt a given fault tolerance mechanism, but to experiment the capabilities of the component-based middleware to manipulate a component-based software architecture at runtime. To this aim, we target the above-described component-based implementation of the PBR strategy. In our scenario, the implementation of the recovery, i.e., when backup becomes the primary due to the primary failing, is not built-in; we show that this recovery procedure can be done by manipulating  components on-the-fly. This can be regarded as a dynamic reconfiguration in the small, involving a single mechanism.
The performed actions are the following:
\begin{itemize}
\item the client is stopped;
\item a reference to the requested operation is obtained by introspecting the client component;
\item the binding between the client and the former primary is disconnected;
\item a new connection between the client and the backup is established
\end{itemize}

\begin{figure}[!ht]
	\begin{framed}
\DefineVerbatimEnvironment{code}{Verbatim}{fontsize=\footnotesize}
\begin{code}
action switchServer(){
root = $domain/scachild::pbr;
c = $root/scachild::client_machine;  
s1 = $root/scachild::primary;
s2 = $root/scachild::backup;
c-ref = $c/scareference::computeService;
s1-serv = $s1/scaservice::computeService;    
s2-serv = $s2/scaservice::computeService;
set-state($c, "STOPPED");
remove-scawire($c-ref,$s1-serv);	
add-scawire($c-ref,$s2-serv);
set-state($c, "STARTED"); }
\end{code}
\end{framed}
\caption{The script for reconfiguring our component-based PBR}

\label{fig:script}
\end{figure}
The script in Figure~\ref{fig:script} implements the previously defined actions. This script can be executed step-by-step using the the OW2 FraSCAti interactive Explorer (a sort of testing phase), but can also be directly executed.
\section{Overall Development Process and Future Works}
Through this experimental implementation, we explored the functionalities of OW2 FraSCAti and assessed its suitability for our resilient computing framework. 

The work described in this paper is part of a development process aiming to provide both a methodology and a tool-box for building resilient systems. The overall process is described in \cite{springerlink:serene}. The main steps identified in this process are: 
\begin{itemize}
\item building a classification of FTMs based on parameters such as the ones enumerated in section \ref{problem};
\item designing FTMs for adaptation, i.e. foreseeing future transitions between them and possible combinations;
\item mapping these designs on components and bindings;
\item writing and executing scripts performing the desired transitions at runtime, in response to different changes in the environment and in the application.
\end{itemize}
A typical scenario to validate our approach can be the following: at initial time, a duplex strategy was selected for a given function to comply with the crash fault model, as requested in the specification; at a later time,  the monitoring of the system reveals a high number of transient physical faults impacting the results of the function. It is decided to combine a time redundancy strategy with the duplex strategy to comply with this new situation. This involves manipulating the component architecture of a duplex strategy to insert components performing time redundancy. Our current experiments show that the facilities provided by the component-based middleware enable this reconfiguration to be done at runtime.

\section{Conclusion}
Our approach to resilient computing relies on several steps, from a design for adaptation down to the verification of the consistency of distributed updates of a real system. A cornerstone in this process is the manipulation of software architecture at runtime. A component-based middleware is an essential piece of this process. The simple experiments briefly described in this paper show that the provided observation and control capabilities enable resilient computing to be addressed.



\bibliographystyle{IEEEtran}
\bibliography{IEEEabrv,edcc}
%

\end{document}